\documentclass[aps,prl,reprint,superscriptaddress,amsmath,amssymb]{revtex4-1}

\usepackage{bm}
\usepackage{braket}
\usepackage{graphicx}

\renewcommand{\vec}[1]{\bm{#1}}
\newcommand{\fermion}{\hat{\psi}}
\newcommand{\boson}{\hat{\phi}}
\newcommand{\veck}{\vec{k}}
\newcommand{\vecp}{\vec{p}}
\newcommand{\vecq}{\vec{q}}
\newcommand{\vecr}{\vec{r}}

\newcommand{\delete}[1]{}

\begin{document}

\title{$P$-Wave Contact Tensor -- Universal Properties
of Axisymmetry-Broken $P$-Wave Fermi Gases}

\author{Shuhei M. Yoshida}
\affiliation{Department of Physics, University of Tokyo,
7-3-1 Hongo, Bunkyo-ku, Tokyo 113-0033, Japan}
\author{Masahito Ueda}
\affiliation{Department of Physics, University of Tokyo,
7-3-1 Hongo, Bunkyo-ku, Tokyo 113-0033, Japan}
\affiliation{RIKEN Center for Emergent Matter Science (CEMS),
Wako, Saitama 351-0198, Japan}

\date{\today}

\begin{abstract}
We investigate universal properties of a $p$-wave Fermi gas with a resonant
interaction in which the axisymmetry is broken spontaneously or externally.
Here, the short-range correlations can be completely characterized by
the nine-component $p$-wave contact tensor, which can be measured by applying
a generalized adiabatic sweep theorem.
The distinctive features of the $p$-wave contact tensor emerge
in a normal $p$-wave Fermi gas in an anisotropic trap and in a superfluid phase.
 An experimental scheme to measure the $p$-wave contact tensor and
 test the adiabatic sweep theorem is also discussed.
\end{abstract}


\maketitle

Universal properties in ultracold atomic gases with resonant interactions
have been gathering great interest.
Among such universal properties are universal relations in the BCS-BEC crossover
~\cite{Tan2008a,*Tan2008b,*Tan2008c,Braaten2008a,Werner2009,Zhang2009,
Combescot2009a}, which predict universal power laws in the short-range
correlations such as a high-momentum asymptote of the momentum distribution and
relate them to macroscopic properties such as a thermodynamic function.
They hold in the resonant regime at any temperature,
in the normal or superfluid phase, and in a trapped or uniform system.
Here, a single quantity called Tan's contact characterizes both short- and
long-distance properties of the system. These theoretical predictions have also
been also verified experimentally~\cite{Stewart2010,Kuhnle2010,Sagi2012}.

Recently, such universal relations have been investigated in a Fermi gas with
a $p$-wave Feshbach resonance;
the notion of the contact has been extended~\cite{Yoshida2015a,Yu2015a,He2016}
and experimentally measured~\cite{Luciuk2015}.
A new feature that the $p$-wave interaction introduces is anisotropy;
a magnetic field that controls the Feshbach resonance lifts the three-fold
degeneracy of the $p$-wave resonance into two distinct ones via the magnetic
dipole-dipole interaction~\cite{Ticknor2004}.
This can make system's correlation functions anisotropic,
and the three-component $p$-wave contact has been introduced
in Refs.~\cite{Yu2015a,He2016}, which is justified when the system possesses
the axisymmetry.
However, breaking of the axial symmetry often plays an crutial role
in this system.
For example, the axial symmetry can be spontaneously broken in a $p+i\beta p$
superfluid phase
~\cite{[][{. See also arXiv:cond-mat/0410620v3 for a correction.}]Gurarie2005,
Cheng2005}.
Another example is a $p$-wave gas confined in an anisotropic trap,
in which external fields may break the symmetry.
Such a geometry has been utilized to investigate phenomena such as
a confinement-induced resonance~\cite{Gunter2005,Granger2004}.

\begin{figure}
	\includegraphics[width=7cm]{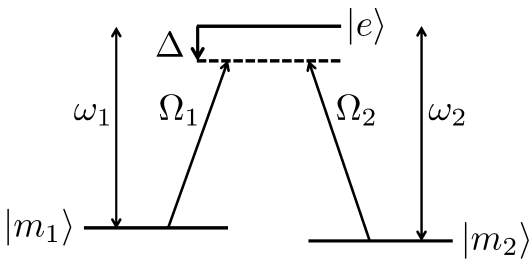}
	\caption{$\Lambda$ configuration among $\ket{m_1}$, $\ket{m_2}$,
	and $\ket{e}$. Here, $\omega_i$ ($i=1,2$) is the resonance frequency between
	$\ket{m_i}$ and $\ket{e}$, and $\Omega_i$ is the complex Rabi frequency
	which is a complex number.
	The two closed-channel states are assumed to be at the two-photon resonance
	detuned by $\Delta$ from $\ket{e}$.\label{fig:lambda}}
\end{figure}

The purpose of this Letter is to point out that, once we allow the axisymmetry
breaking, a complete characterization of the short-range correlations requires
the nine-component $p$-wave contact tensor $C_{m,m'}$
~\footnote{The notion of the $p$-wave contact tensor was first presented at
Few-body Physics in Cold Atomic Gases (April 14-15, 2016, Beijing) and
the 47th Regular DAMOP Meeting (May 25, 2016, Providence).}.
For example, the momentum distribution behaves as
\begin{align}
	n_{\vec{k}}
	\sim k^{-2} \sum_{m,m'=-1}^1 C_{m,m'}
		Y_1^{m\ast}(\hat{\vec{k}}) Y_1^{m'}(\hat{\vec{k}})
	\label{eq:def-contact-tensor}
\end{align}
for $|\vec{k}| \gg k_F, \lambda_T$, where $k_F$ is the Fermi momentum,
and $\lambda_T$ is the thermal de Broglie length.
This implies that $C_{m,m'}$ contains the three-component $p$-wave contact
as the three diagonal components.
We show the following generalized adiabatic sweep theorem:
\begin{align}
	C_{m,m'} = \frac{32\pi^2M}{\hbar^2} \frac{\partial E}{\partial(-1/v_{m,m'})},
	\label{eq:adiabatic sweep theorem}
\end{align}
where $v_{m,m'}$ defined later determines the low-energy scattering phase shift
of transitions between the channels with the projections of
the relative angular momentum $m$ and $m'$.
The diagonal component $v_{m,m}$ is the usual $m$-dependent $p$-wave scattering
volume which can be controlled by a $p$-wave Feshbach resonance.
On the other hand, $v_{m,m'}$ with $m\neq m'$ is associated with
an unconventional $p$-wave scattering that does not conserve the projection of
the relative angular momentum of two colliding atoms, the consequences of which
have not been investigated yet.
There are no such processes in ordinary $p$-wave Fermi gases because
the $p$-wave Feshbach resonance is anisotropic but still axially symmetric.
However, we show that a Raman process in the $\Lambda$ scheme
(see Fig.~\ref{fig:lambda}) can be used to control
the unconventional $p$-wave scattering, test the adiabatic sweep theorem and
measure the entire $p$-wave contact tensor.
We also demonstrate that the off-diagonal components of $C_{m,m'}$ emerge
in the $p+i\beta p$ superfluid phase and a normal Fermi gas
in a pancake-shaped trap.

First, let us discuss why the $p$-wave contact has to be promoted to a tensor.
To be specific, we base our discussion on the following two-channel model of
a single-component Fermi gas
with a resonant $p$-wave interaction~\cite{Yoshida2015a}:
\begin{equation} \begin{split}
\hat{H} &= \int d^3\vec{r} \,
	\frac12 |\nabla \fermion(\vec{r}) |^2 \\
&+ \int d^3 \vec{r} \left[
		\frac14 |\nabla\boson_m (\vec{r})|^2
		+ \epsilon_{m,m'} \boson_m^\dagger \boson_{m'}(\vec{r})
	\right] \\
& + \int d^3\vec{r}_1 d^3\vec{r}_2 \,\left[
	\frac{g}2 u_m( \vec{r}_{12} )
		\fermion^\dagger (\vec{r}_1) \fermion^\dagger(\vec{r}_2)
		\boson_m(\vec{R}_{12}) \right. \\
&	\left.	+ \frac{g^\ast}{2} u_m^\ast( \vec{r}_{12} )
		\boson_m^\dagger(\vec{R}_{12})
		\fermion(\vec{r}') \fermion (\vec{r})
	\right],
\label{eq:hamiltonian}
\end{split} \end{equation}
where summation over $m,m'=-1,0,1$ is implied.
Here $\fermion$ is a fermionic field operator representing the open-channel
atoms, $\boson_m$ is a bosonic operator representing the closed-channel
molecules, $\vec{r}_{12}\equiv\vec{r}_1-\vec{r}_2$, and
$\vec{R}_{12}\equiv\frac{\vec{r}_1+\vec{r}_2}{2}$.
We set $\hbar=k_B=M=1$, where $\hbar$ is the Planck constant divided by $2\pi$,
$k_B$ is the Boltzmann constant, and $M$ is the mass of the atom.
The closed-channel energy tensor $\epsilon_{m,m'}$ and a coupling constant $g$
control the Feshbach resonance, and are determined so that
they reproduce the $p$-wave scattering volume and the effective range.
In the presence of a magnetic field parallel to the $z$ axis,
$\epsilon_{m,m'}$ is diagonal and the Hamiltonian reduces to the familiar
expression of the two-channel model~\cite{Gurarie2005}.
In particular, the anisotropy of the $p$-wave Feshbach resonance due to
the magnetic field can be taken into account in the $m$-dependence
of $\epsilon_{m,m}$.
We use the tensor here to make the expression of the Hamiltonian unchanged
under coordinate rotation.
The coupling function $u_m(\vecr)$ regularizes the short-range singularity of
the interaction and is normalized so that
its Fourier transform $\tilde{u}_m(\vec{k})$ is $\sim kY_1^m(\hat{\vec{k}})$
for $k\ll r_0^{-1}$, where $r_0$ is the range of the interaction.

Following Ref.~\cite{Yoshida2015a}, we introduce
a set of wave functions $\{\Psi^{(N_o,N_c)}\} $
characterizing a many-body state.
Each $\Psi^{(N_o,N_c)}  ( \{\vecp\} ; \{ \vecq \}, \{ m \} )$ represents
a state with $N_o$ open-channel atoms and $N_c$ closed-channel molecules,
in which $\{\vecp\}$, $\{ \vecq \}$, and $\{ m \}$ are the sets of
the atomic momentum $\vecp_i$, the molecular momentum $\vecq_i$, and
the angular momentum projection $m_i$ of the molecular rotation, respectively.
If we expand the time-independent Schr\"{o}dinger equation
in terms of $1/p_i$, we obtain
\begin{align} \begin{split}
&\Psi^{(N_o,N_c)} ( \{\vecp\} ; \{ \vecq \}, \{ m \} )
 \sim \sum_{j=1, j\neq i}^{N_o} \sum_{m_{ji}=-1}^1 (-1)^{j+1} g  \\
&    \frac{Y_1^{m_{ij}}(\hat{\vec{p}}_{ij})}{p_{ij}}
    \Psi^{(N_o-2,N_c+1)} ( \{ \vecp \}_{ij}; \{ \vecq \}_{ij}, \{ m \}_{ij} )
\end{split} \label{eq:asymptotic_wf}
\end{align}
to the leading order. Here, $\vecp_{ij} \equiv \frac{\vecp_i-\vecp_j}{2}$,
$\{ \vecp \}_{ij}$ is $\{ \vecp \}$ excluding $\vecp_i$ and
$\vecp_j$, $\{ \vecq \}_{ij}$ is $\{ \vecq \}$ including $\vec{p}_i+\vec{p}_j$,
and $\{ m\}_{ij}$ is $\{m\}$ including $m_{ij}$.
This result indicates that an asymptotic wave function is in general
a linear combination of terms proportional to
the spherical harmonic function $Y_1^m(\hat{p}_{ij})$.
When calculating the momentum distribution, two things should be noted.
First, the momentum distribution is calculated from the squared wave function,
which contains terms like
$Y_1^{m\ast}(\hat{\vec{p}}_{ij}) Y_1^{m'}(\hat{\vec{p}}_{ij})$.
Second, for terms with $m\neq m'$ to vanish, the axisymmetry of the microscopic
interaction is not sufficient, but the state has to be invariant under rotation
around the $z$ axis.
Since we do not make the assumption of the axial invariance here,
the momentum distribution has in general the high-momentum tail
of the form in Eq.~(\ref{eq:def-contact-tensor}) up to the leading order.

We also find from Eq.~(\ref{eq:asymptotic_wf}) the following expression of
$C_{m,m'}$ within our two-channel model:
\begin{align}
C_{m,m'} = |g|^2 \int \frac{d\veck}{(2\pi)^3}
\langle \boson_m^\dagger(\veck) \boson_{m'}(\veck) \rangle.
\label{eq:two_channel_contact}
\end{align}
This expression gives useful interpretation of $C_{m,m'}$
in the ultracold atomic experiments as the number of the closed-channel
molecules in an $m$ state for $m=m'$ and the coherence between two molecular
states $m$ and $m'$ for $m\neq m'$. The $p$-wave contact tensor belongs
to the $3\otimes3$ representation of the rotation group.
Therefore, it can be decomposed into one-, three-, and five-dimensional
irreducible representations, each of which can also be interpreted as
the number, angular momentum, and nematicity of the closed-channel molecules,
respectively.


The $p$-wave contact tensor is directly related to the thermodynamics through
the adiabatic sweep theorem.
To derive the theorem, we need to define the generalized $p$-wave scattering
volume $v_{m,m'}$. If the $p$-wave scattering from an $m$ state into an $m'$
state is allowed, the scattering amplitude takes the following form:
\begin{align}
f(\hat{\vec{p}}, \hat{\vec{p}}', k)
= \sum_{m,m'=-1}^1 4\pi f_{m,m'}(k)
	Y_1^{m}(\hat{\vec{p}}) Y_1^{m'\ast}(\hat{\vec{p}}'),
\end{align}
where $k\hat{\vec{p}}'$ is the incoming momentum and $k\hat{\vec{p}}$ is
the outgoing momentum.
We can then define $v_{m,m'}$ and $k_{m,m'}^\mathrm{(eff)}$ from the inverse of
$f_{m,m'}(k)$ as a $3\times 3$ matrix and its low-energy expansion,
\begin{align}
(f^{-1})_{m,m'}(k)
= -\frac{1}{v_{m,m'}k^2} + \frac{1}{2} k^\mathrm{(eff)}_{m,m'}
	- ik\delta_{m,m'} + O(k^2).
\end{align}
Note that if $f_{m,m'}(k)=0$ for $m\neq m'$, as in ordinary cases,
$1/v_{m,m'}$ rather than $v_{m,m'}$ vanishes.

We are prepared to show the adiabatic sweep theorem
for the $p$-wave contact tensor.
Here, we temporarily remove the assumption that the Feshbach resonance
is axisymmetric; that is, we allow $\epsilon_{m,m'}$ to have an arbitrary form.
Then we can solve the two-body problem and calculate the scattering amplitude
to determine $\epsilon_{m,m'}$ and $g$
in terms of $1/v_{m,m'}$ and $k^\mathrm{(eff)}_{m,m'}$.
We can use those expressions to obtain
\begin{align}
\frac{\partial \hat{H}_{\mathrm{det}}}{\partial (-1/v_{m,m'})}
= \frac{\hbar^2 |g|^2}{32 \pi^2 M}
\int d^3\vec{r} \, \boson_m^\dagger(\vec{r}) \boson_{m'}(\vec{r}).
\end{align}
On the other hand, the Hellmann-Feynman theorem reads
\begin{align}
\frac{\partial E}{\partial (-1/v_{m,m'})} =
\left\langle
    \frac{\partial \hat{H}_{\mathrm{det}}}{\partial (-1/v_{m,m'})}
\right\rangle.
\end{align}
Combining these with Eq.~(\ref{eq:two_channel_contact}),
we obtain the generalized adiabatic sweep theorem~(\ref{eq:adiabatic sweep theorem}).
After taking the derivative with respect to $-1/v_{m,m'}$,
we can set the off-diagonal elements of $\epsilon_{m,m'}$ to be zero
as they should be in realistic Feshbach resonance.
We emphasize that the off-diagonal components of $C_{m,m'}$ can be nonzero,
in general,
even in the absence of the off-diagonal components of $\epsilon_{m,m'}$.

No Feshbach resonance can tune $1/v_{m,m'}$ for $m\neq m'$.
Here we show that it can be controlled by using a two-photon Raman process.
Figure~\ref{fig:lambda} shows the proposed configuration of the molecular
levels and lasers.
Let $\ket{e}$ be a diatomic molecular state that couples to the closed-channel
states $\ket{m_1}$ and $\ket{m_2}$ via lasers of appropriate frequencies and
polarizations.
We denote the resonant frequencies for $\ket{m_1}$ and $\ket{m_2}$
by $\omega_1$ and $\omega_2$, respectively.
With the rotating wave approximation, the effective Hamiltonian density
$\mathcal{H}_\mathrm{Raman}$ of this three-level system is
\begin{align}
	\hat{\mathcal{H}}_\mathrm{Raman}
&=
	\left(
	\begin{array}{ccc}
		\hat{\phi}_{m_1}^\dagger
		& \hat{\phi}_{m_2}^\dagger
		& \hat{\phi}_{e}^\dagger
	\end{array}
	\right) h_\mathrm{Raman}
		\left(
		\begin{array}{c}
			\hat{\phi}_{m_1} \\ \hat{\phi}_{m_2} \\ \hat{\phi}_{e}
		\end{array}
		\right), \\
	h_\mathrm{Raman}
&=
		\left(
		\begin{array}{ccc}
			0 & 0 & \Omega_1^\ast/2 \\
			0 & 0 & \Omega_2^\ast/2 \\
			\Omega_1/2 & \Omega_2/2 & \Delta
		\end{array}
		\right) ,
\end{align}
where $\hat{\phi}_e$ is the bosonic annihilation operator of molecules
in $\ket{e}$, $\Omega_i$ ($i=1,2$) is the complex Rabi frequency, and
$\Delta$ is the detuning.
For the moment, the spatial argument is omitted from
the bosonic annihilation operators.
The eigenvalues of $h_\mathrm{Raman}$ is
$0, \frac12(\Delta\pm\sqrt{\Delta^2 + |\Omega_1|^2+|\Omega_2|^2})$,
which we denote by $\omega_0$ and $\omega_\pm$ and
the corresponding molecular states by $\ket{\bar{0}}$ and $\ket{\pm}$.
The eigenvectors of $h_\mathrm{Raman}$ are, apart from the normalization,
$(\Omega_2, -\Omega_1, 0)^T$ and $(\Omega_1^\ast,\Omega_2^\ast,2\omega_\pm)^T$
for $\omega_0$ and $\omega_\pm$, respectively.
We can rewrite $\hat{\mathcal{H}}_\mathrm{Raman}$ as
\begin{align}
	\hat{\mathcal{H}}_\mathrm{Raman}
	= \omega_0 \hat{\phi}_0^\dagger \hat{\phi}_0
		+ \omega_- \hat{\phi}_-^\dagger \hat{\phi}_-
		+ \omega_+ \hat{\phi}_+^\dagger \hat{\phi}_+,
	\label{eq:diag_Raman}
\end{align}
where $\hat{\phi}_0$ and $\hat{\phi}_\pm$ are the bosonic field operators
corresponding to $\ket{\bar{0}}$ and $\ket{\pm}$.
Among them, $\hat{\phi}_0$ and $\hat{\phi}_-$ are adiabatically connected
to $\hat{\phi}_{m_1}$ and $\hat{\phi}_{m_2}$ in the weak field limit.
Now, suppose that no molecules are in $\ket{e}$ at the initial time and
that the intensities of the lasers are adiabatically ramped up.
Then we can ignore the third term in Eq.~(\ref{eq:diag_Raman}),
and for weak laser fields such that
$|\Omega_1| \ll \Delta$ and $|\Omega_2| \ll \Delta$, we obtain
\begin{align}
	\begin{split}
	\hat{\mathcal{H}}_\mathrm{Raman}
&	\simeq \frac{1}{4\Delta}
		\left( |\Omega_1|^2 \hat{\phi}_{m_1}^\dagger \hat\phi_{m_1}
			+ |\Omega_2|^2 \hat\phi_{m_2}^\dagger \hat\phi_{m_2} \right. \\
& \left. \quad
			+ \Omega_2^\ast \Omega_1 \hat\phi_{m_1}^\dagger \hat\phi_{m_2}
			+ \Omega_1^\ast \Omega_2
				\hat\phi_{m_2}^\dagger \hat\phi_{m_1} \right).
	\end{split}
\end{align}
This amounts to adiabatically sweeping
$\epsilon_{m_1,m_2} = \epsilon_{m_2,m_1}^\ast$ as
$\Omega_2^\ast \Omega_1/4\Delta$, and
thus to tuning $1/v_{m_1,m_2}$ for $m_1\neq m_2$.

The same configuration can be used to measure
the off-diagonal components of $C_{m,m'}$.
This time, the lasers are suddenly turned on and the frequencies are set
close to the resonance $\Delta \ll |\Omega_1|, |\Omega_2|$.
If the lifetime of the excited state $\ket{e}$ is much shorter than
that of $\ket{m_1}$ and $\ket{m_2}$, there remain only the molecules
in $\ket{\bar{0}}$ after some time because $\ket{\bar{0}}$,
so called the dark state, does not contain $\ket{e}$.
Therefore, by counting the number of the molecules,
we obtain $\langle \hat{\phi}_0^\dagger \hat{\phi}_0\rangle
= \langle \frac{1}{|\Omega_1|^2+|\Omega_2|^2}
|\Omega_2\hat{\phi}_1-\Omega_1\hat{\phi}_2|^2 \rangle$,
and by repeating the measurement with the different amplitudes and
the relative phases of $\Omega_1$ and $\Omega_2$, we can determine
$\langle \hat{\phi}^\dagger_{m} \hat{\phi}_{m'}\rangle$ for all $m$ and $m'$.
Once we know $\langle \hat{\phi}^\dagger_{m} \hat{\phi}_{m'}\rangle$,
the $p$-wave contact tensor is determined by Eq.~(\ref{eq:two_channel_contact}).

In the remaining part, we discuss possible physical situations in which
the off-diagonal components of $C_{m,m'}$ are significant in systems
with $1/v_{m,m'}=0$ for $m\neq m'$.
Specifically, we take two examples, a $p$-wave superfluid and
a normal Fermi gas in an anisotropic trap.

A $p$-wave superfluid provides an example which can exhibit the off-diagonal
components of the $p$-wave contact tensor due to the spontaneous
axisymmetry breaking.
If we perform the mean field approximation to the Hamiltonian
(\ref{eq:hamiltonian})
with the order parameter~\cite{Cheng2005,Gurarie2005,Ohashi2005}
$B_m \equiv \frac{1}{\sqrt{V}} \langle \hat{b}_{m\vec{0}} \rangle$,
we can calculate the momentum distribution as
\begin{align}
n_{\vec{p}} \sim \frac{|g|^2}{4 p^2} \left| \sum_{m=-1}^1 Y_{1}^m(\hat{\vec{p}}) B_m \right|^2 .
\end{align}
From this expression, we can see that the $p$-wave contact tensor has
the following mean-field contribution:
\begin{align}
C_{m,m'}^\mathrm{MF} = \frac{|g|^2}{4} B_m B_{m'}^\ast.
\end{align}
This takes the diagonal form if and only if the system is in the $p$ or $p+ip$
superfluid phase.
In the presence of moderate anisotropy of the $p$-wave Feshbach resonance,
it has been shown that a Fermi gas described by the Hamiltonian
(\ref{eq:hamiltonian}) undergoes the spontaneous breaking of the axial symmetry
in superfluid phase and a phase transition into the $p+i\beta p$ superfluid,
which is neither the $p$ nor $p+ip$ superfluid~\cite{Gurarie2005,Cheng2005}.
Therefore, the off-diagonal components of $C_{m,m'}$ appear in the $p$-wave
superfluid with an anisotropic Feshbach resonance, and the emergence of
the off-diagonal components signals the transition from the $p$ to
the $p+i\beta p$ superfluid.


A normal Fermi gas in an anisotropic trap may also reveal the off-diagonal
$p$-wave contact, at least from the symmetry argument.
On the other hand, to the extent that
the local-density approximation (LDA) is valid,
the off-diagonal components may be zero or very small because
the assumption of the local uniformity in the LDA implies
that the $p$-wave contact tensor density is diagonal at each point in space.
Therefore, the off-diagonal $p$-wave contact tensor in the normal phase is
an indication that the LDA breaks down.

\begin{figure}
	\includegraphics[width=8.6cm]{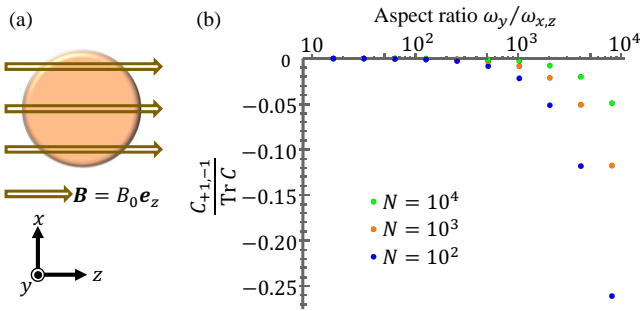}
	\caption{(Color online) Off-diagonal component of the $p$-wave contact
	tensor in a Fermi gas in a pancake-shaped trap.
	(a) Configuration of the trap and magnetic field.
	The gas is tightly confined in the $y$ direction.
	(b) Calculated $C_{+1,-1}$ normalized by $\mathrm{Tr}\,C$ with
	several aspect ratio at $T=2T_F$.
	We use the interaction parameters of the $m=\pm1$ resonance of
	the $\ket{F=9/2,m_F=-7/2}$ state of ${}^{40}\mathrm{K}$ at
	$B_0=198.3\,\mathrm{G}$.
	The number of particles is $10^2$ (blue dots), $10^3$ (orange dots),
	and $10^4$ (green dots). \label{fig:pancake}}
\end{figure}
To examine what happens in reality, we calculate the $p$-wave contact tensor up
to the second order in the cluster expansion~\cite{Huang1987}.
Specifically, we consider a Fermi gas in a pancake-shaped harmonic trap
with $\omega_y > \omega_x=\omega_z =100\,\mathrm{Hz}$ and
with an external magnetic field parallel to the $z$ axis
to control the Feshbach resonance, as depicted in Fig.~\ref{fig:pancake}(a).
The temperature is fixed at $T=2T_F$. We take the interaction parameters for
the Feshbach resonance at $B=198.3\,\mathrm{G}$ in the $m=\pm1$ channel
in the $\ket{F=9/2, m_F=-7/2}$ state of ${}^{40}\mathrm{K}$~\cite{Ticknor2004}.
We assume that the atoms are non-interacting in the $m=0$ channel
to avoid a large contribution from the $m=0$ dimer;
this assumption is consistent with the experimental observation that
the $p$-wave contact does not grow on the BEC side of the Feshbach resonance
on the experimental time scale~\cite{Luciuk2015}.

Figure~\ref{fig:pancake}(b) shows the calculated $C_{+1,-1}=C_{-1,+1}^\ast$
normalized by the trace of the $p$-wave contact tensor $\mathrm{Tr}\,C$
with the varying aspect ratio and the total number of atoms including the atoms
in both open and closed channels. The other off-diagonal components are
zero due to the symmetry of the configuration.
The minus sign implies that the correlation
in the tightly confined $y$ direction is weaker than that in the $x$ direction.
For the small aspect ratio, $C_{+1,-1}$ is vanishingly small.
Indeed, the tighter trap frequency $\omega_y$ exceeds the temperature $T$
at the aspect ratio $\sim 70$ for $N=10^2$ and $\sim 700$ for $N=10^4$;
only for the higher aspect ratio than these does the growth of $C_{+1,-1}$
become visible.
This implies that at this temperature, the quasi-two-dimensional regime should
exhibit a significant amount of $C_{+1,-1}$. Note that the ratio
$C_{+1,-1}/\mathrm{Tr}\,C$ depends on $N$ only through the temperature
$T=2(\omega_x\omega_y\omega_z 6N)^{1/3}$ in this approximation,
and that the smaller $N$ corresponds to the lower temperature.
Thus one may infer that a larger $C_{+1,-1}/\mathrm{Tr}\,C$ can be obtained
in the current experimental regime of $T\sim 0.2T_F$.
At that low temperature, however, the current approximation breaks down,
and another method is needed.

%

In conclusion, we have discussed the universal properties of a Fermi gas with
a resonant interaction. We have introduced the $p$-wave contact tensor,
which characterizes both short-range correlations and thermodynamics
of such a system.
We have generalized the adiabatic sweep theorem applicable
to all of the nine components of the $p$-wave contact tensor.
As the theorem is stated in terms of the parameters characterizing
the unconventional type of the $p$-wave interaction,
we have proposed a Raman process in the $\Lambda$ scheme to manipulate
these parameters and to measure the $p$-wave contact tensor.
We have also investigated the $p$-wave contact tensor in the $p$-wave
superfluid and a $p$-wave Fermi gas in a pancake-shaped trap,
where the axial symmetry is broken either spontaneously or externally.
In the superfluid phase, the appearance of the off-diagonal component implies
the predicted transition from the $p$ phase to the $p+i\beta p$ phase.
We have argued the possibility that due to the beyond-LDA effect,
the off-diagonal components can be observed in an anisotropic trap,
even in the normal phase.

This work was supported by
JSPS Grants-in-Aid for Scientific Research (KAKENHI Grant No.~JP26287088),
MEXT Grant-in-Aid for Scientific Research on Innovative Areas
``Topological Materials Science'' (KAKENHI Grant No.~JP15H05855),
the Photon Frontier Network Program from MEXT of Japan,
and the Mitsubishi Foundation.
SMY was supported by Grant-in-Aid for JSPS Fellows
(KAKENHI Grant No.~JP16J06706),
and the Japan Society for the Promotion of Science through Program
for Leading Graduate Schools (ALPS).

{\em Note added}. After completion of this work, we became aware of
a closely related work by Zhang, He, and Zhou~\cite{Zhang2016},
in which they take into account the scattering in all partial waves.
Our conclusions agree where overlapping.

\bibliography{library}

\end{document}